\documentclass[%
 letterpaper,
superscriptaddress,
 amsmath,amssymb,
 aps,
]{revtex4-2}

\usepackage[utf8]{inputenc}
\usepackage[T1]{fontenc}

\usepackage{braket}
\usepackage{graphicx}
\usepackage{dcolumn}
\usepackage{bm}
\usepackage{lipsum}
\usepackage{tikz}
\usetikzlibrary{quantikz}
\usepackage{adjustbox}
\usepackage{subfigure}
\usepackage{microtype}
\usepackage{todonotes}
\usepackage{setspace}
\usepackage{hyperref}
\usepackage{soul}
\usepackage{amsmath}
\usepackage{amssymb}
\usepackage{amsfonts}

\begin{document}

\title{Using Differential Evolution to avoid local minima  
in Variational Quantum Algorithms}

\author{Daniel Failde}
\affiliation{Centro de Supercomputación de Galicia (CESGA), 15705 Santiago de Compostela, Spain}
\author{Jose Daniel Viqueira}
\affiliation{Centro de Supercomputación de Galicia (CESGA), 15705 Santiago de Compostela, Spain}
\affiliation{Computer Graphics and Data Engineering (COGRADE), Departamento de Electrónica e Computación, Universidade de Santiago de Compostela, 15782 Santiago de Compostela, Spain}
\author{Mariamo Mussa Juane}
\affiliation{Centro de Supercomputación de Galicia (CESGA), 15705 Santiago de Compostela, Spain}
\author{Andres Gomez }
\affiliation{Centro de Supercomputación de Galicia (CESGA), 15705 Santiago de Compostela, Spain}

\email[]{dfailde@cesga.es}


\begin{abstract}
Variational Quantum Algorithms (VQAs) are among the most promising NISQ-era algorithms for harnessing quantum computing in diverse fields. However, the underlying optimization processes within these algorithms usually deal with local minima and barren plateau problems, preventing them from scaling efficiently. Our goal in this paper is to study alternative optimization methods that can avoid or reduce the effect of these problems. To this end, we propose to apply the Differential Evolution (DE) algorithm to VQAs optimizations. Our hypothesis is that DE is resilient to vanishing gradients and local minima for two main reasons: (i) it does not depend on gradients, and (ii) its mutation and recombination schemes allow DE to continue evolving even in these cases. To demonstrate the performance of our approach, first, we use a robust local minima problem to compare state-of-the-art local optimizers (SLSQP, COBYLA, L-BFGS-B and SPSA) against DE using the Variational Quantum Eigensolver algorithm. Our results show that DE always outperforms local optimizers. In particular, in exact simulations of a 1D Ising chain with 14 qubits, DE achieves the ground state with a 100\% success rate, while local optimizers only exhibit around 40\%. We also show that combining DE with local optimizers increases the accuracy of the energy estimation once avoiding local minima. Finally, we  demonstrate how our results can be extended to more complex problems by studying DE performance in a 1D Hubbard model.

\end{abstract}


\flushbottom
\maketitle                     


\section{Introduction}
Variational Quantum Algorithms (VQAs) are a promising group of hybrid classical-quantum algorithms that can reach \textit{quantum advantage} for solving many relevant problems in NISQ-era quantum computers \cite{Cerezo2021}. VQAs are composed of classical and quantum routines, which are used to minimize a given cost function. They use (i) quantum computing to evaluate a  function with a parameterized quantum circuit, and (ii) classical computing to optimize the circuit parameters on each iteration. Their parameterized structure gives VQAs the flexibility to work with shallow quantum circuits, unlike other quantum algorithms such as the Quantum Phase Estimation, Shor\textquotesingle s and Grover\textquotesingle s algorithms \cite{nielsen,Shor,Grover}. Therefore, VQAs can circumvent the issues associated with the errors and coherence times intrinsic to current quantum hardware to generate reliable outputs even without error correction techniques.   In addition, their general purpose (minimizing a function) allows using them for many different optimization problems in machine learning, chemistry, physics, and mathematics, among others \cite{ RevModPhys.92.015003,TILLY20221,PhysRevA.101.010301,PhysRevA.99.062304,9144562}. One of the most common algorithms in the family of VQAs is the Variational Quantum Eigensolver (VQE), which aims to find the quantum state (or the set) that minimizes the energy of a given Hamiltonian \cite{TILLY20221,Fedorov2022}. On a small scale, VQE can solve, for instance, small molecules and lattice models \cite{Kandala2017,PhysRevResearch.4.013165}. But on a broader framework, it is an excellent candidate to simulate large chemical reactions, perform exact calculations on crystalline solids, and uncover the physics behind complex systems such as the Hubbard model or exotic states of matter \cite{PhysRevLett.79.2586,PhysRevA.92.062318,Stanisic2022,PhysRevResearch.3.013184,PhysRevLett.121.110504,PhysRevApplied.9.044036}. 

However, currently, VQE and VQAs have scaling problems \cite{Anschuetz2022}. Typical issues concern ansatz expressibility and trainability, which refers to the degree of information that a quantum circuit has to reproduce an energy state of the system and the easiness of fitting the parameters to find the global minimum, respectively \cite{PhysRevLett.128.080506,PRXQuantum.3.010313}. Both concepts are directly related to problems in the optimization landscape where VQAs can present many local minima and barren plateaus \cite{PRXQuantum.3.010313,Romero_2019,PhysRevLett.127.120502, Anschuetz2022,McClean2018}. These problems worsen with the number of qubits since the Hilbert space grows exponentially \cite{Anschuetz2022,McClean2018}. Local minima inherently arise from minimizing a complex function. Barren plateaus are flat areas in the cost function landscape of VQAs where gradients vanish exponentially with the problem size  \cite{McClean2018}. These areas can arise from different sources, such as random parameterized  quantum circuits, noisy environments, and a high degree of entanglement \cite{Cerezo_barren,Arrasmith2021effectofbarren}. Vanishing gradients and local minima are severe problems towards scaling VQAs. These cause unsuccessful optimizations as well as a significant increase in the number of measurements needed to estimate tiny gradients \cite{McClean2018, Romero_2019}. Thus, VQE circuit construction requires a smart and educated ansatz selection taking, for example,  some knowledge from your Hamiltonian to reduce the number of parameters to optimize without losing expressibility \cite{Fedorov2022,PhysRevResearch.2.043246}. Additionally, it is important to deploy a suitable optimization method that maximizes the probability of avoiding traps in the optimization landscape and the consequences of dealing with tiny gradients.

In this work, we focus on the optimization problem motivated by the lack of optimization methods that can successfully avoid these issues in VQAs \cite{PhysRevLett.127.120502}.  In fact, there is evidence that usual gradient-based and some gradient-free local optimizers suffer from barren plateaus and local minima syndrome \cite{Arrasmith2021effectofbarren,PhysRevLett.127.120502}. On the other hand, there are recent alternatives that use the Quantum Fisher Information Matrix (QFIM) to lead optimization \cite{Stokes2020quantumnatural, Gacon2021simultaneous}. In particular, the Quantum Natural Gradient has successfully  found the ground state (GS), i.e. the state with minimum energy, of some specific models up to a considerable number of qubits \cite{PhysRevResearch.2.043246}. Gradient-based optimizers using the QFIM are promising methods. However, they still depend on gradients, and in general, they are computationally expensive and imply a significant increase in the number of function evaluations \cite{Stokes2020quantumnatural}. Alternatives designed to decrease the computational complexity, such as QN-SPSA, could not achieve the same performance \cite{Gacon2021simultaneous}. In this article, we analyze an alternative optimization strategy based on Differential Evolution (DE) algorithm. DE is an evolutionary algorithm based on population breeding that is gradient-free, easy to implement, and to parallelize. We expect DE to avoid or drastically reduce the effects of vanishing gradients and local minima since its parameter update can naturally keep evolving even in these cases. 

To test this approach, we  compare DE with four state-of-the-art local optimizers (SLSQP, COBYLA, L-BFGS-B, and SPSA) that have demonstrated good results in the current literature \cite{Kandala2017,PhysRevResearch.4.013165,PRXQuantum.3.010309,McCaskey2019,Sherbert2022,Romero_2019}. For this, we choose a 1D Ising model without a magnetic field. That is a simple model but a good test for optimization. In fact, as we show later, usual local gradient-based and gradient-free optimizers tend to meaningfully fail as one increases the number of qubits/lattice sites in the system. The reasons are an enlarged number of excited states and their growing degeneracy in the parameter space that defines a robust local minima problem for optimization. Our results show how DE can avoid or drastically reduce this problem in the optimization landscape, substantially improving the success rate. We test several simple variations within this genetic algorithm that always outperform local optimization methods for large systems. Specifically, we identify one recombination strategy of DE with an exponential crossover that avoids all local minima in the range studied and is suitable to work together with gradient-based methods to speed up optimization and its convergence. Finally, we test DE performance in a correlated fermionic system by applying the same methodology to a 1D Hubbard model with eight qubits. We show again how DE outperforms local optimizers, indicating its potential usage to scale up the study of strongly correlated systems in quantum devices.

\section{Methods}

\subsection*{Ising model without magnetic field}
The Ising model is one of the most simple and well-studied models in the literature, which serves as a starting point towards more complex models for studying magnetism and describing phase transitions \cite{PFEUTY197079,PhysRev.87.410}. Furthermore, classical optimization problems can also be tackled by mapping them to spin Hamiltonians \cite{Lucas}. An example of this is in the Quantum Approximate Optimization Algorithm (QAOA) \cite{QAOA,Farhi2022quantumapproximate}.  In quantum computing, spin models can also help to get some intuition for circuit construction.  This is because some ansatzes, such as the family of Unitary Coupled Cluster (UCC), QAOA, and Hamiltonian Variational Ansatz (HVA), employ a Hamiltonian to build the parameterized quantum circuit \cite{Fedorov2022,Grimsley2023,PhysRevA.92.042303,PRXQuantum.1.020319}. Thus, working with spin models can also serve to estimate the system connectivity and the degree of entanglement required to simulate our system efficiently \cite{Chen:22}. In its more general form, the Ising model consists of a Hamiltonian 
\begin{equation}
H=-\sum_{\left\langle i,j \right\rangle}^n J_{ij} \sigma_i \sigma_j - \sum_i^n h_i \sigma_i
\end{equation}
where the first summation stands for the interaction between adjacent spins $\sigma \in \{-1,+1\}$, $J_{ij}$ represents the coupling constant between spins in sites $i,j$ and $n$ is the number of sites (qubits) in the lattice (circuit). The second summation represents the coupling of individual magnetic moments with an external magnetic field, where $h_i$ is the magnetic field at site $i$. In the Ising model, each site has two possible values $\pm 1$ so that, in a quantum mechanical description, $\sigma$ can be any of the Pauli matrices. 

Focused on the 1D case, we set the magnetic field $h=0$, maximizing the degeneracy of the excited levels, and the interaction constant $J=J_{ij}=1, \forall i,j$. So, our Hamiltonian in matrix form is a sum of Pauli strings
\begin{equation}
    H =-\sigma \otimes \sigma \otimes \mathbb{I} \otimes \mathbb{I} \otimes ... \otimes \mathbb{I}  
     - \mathbb{I} \otimes \sigma \otimes \sigma \otimes \mathbb{I} \otimes  ... \otimes \mathbb{I} 
    - ...  
    - \mathbb{I} \otimes ... \otimes \mathbb{I} \otimes \mathbb{I} \otimes \sigma \otimes \sigma 
\label{matrix ham}
\end{equation}
being $\mathbb{I}$ the 2$\times$2 identity matrix. $\otimes$ stands for the Kronecker product. We take, without loss of generality, $\sigma=\sigma_y$ and open boundary conditions, so we have a chain of spins with no interaction between the first and last elements ($J_{1n}=0$). In this way, it is straightforward to see that there are just two configurations minimizing the energy $\braket{H}$ with all spins oriented in the same direction. In the present case, these are $\ket{i}^{\otimes n}$, $\ket{-i}^{\otimes n}$ or their superposition, being  $\ket{\pm i}$ the eigenstates of $\sigma_y$. The degeneracy of states in the first excited energy level is $2(n-1)$, and for $n>2$, the second excited level is $(n-1)(n-2)$-fold degenerate. Eigenvalues go from $-(n-1)$ to $n-1$ in steps of two. 

\begin{figure}[t]
    \centering
    \begin{adjustbox}{width=0.6\textwidth}
    \begin{quantikz}
    & \lstick{$\ket{0}$} & \gate{R_y(\pi/4)} & \gate{R_y(\theta_{1,l})}\gategroup[wires=3,steps=4,style={dashed,
    rounded corners,fill=blue!20, inner xsep=2pt},background,label style={label position=right,anchor=north,yshift=2.0cm, xshift=0.25cm}]{{\sc $\times L$} } & \gate{R_z(\theta_{4,l})} & \ctrl{1} & \qw & \gate{R_y(\theta_{1,L+1})}  & \gate{R_z(\theta_{4,L+1})}  \\
     & \lstick{$\ket{0}$} & \gate{R_y(\pi/4)} & \gate{R_y(\theta_{2,l})} & \gate{R_z(\theta_{5,l})} & \ctrl{-1}  & \ctrl{1} & \gate{R_y(\theta_{2,L+1})}  & \gate{R_z(\theta_{5,L+1})} \\
     & \lstick{$\ket{0}$} & \gate{R_y(\pi/4)} &\gate{R_y(\theta_{3,l})} & \gate{R_z(\theta_{6,l})} & \qw  & \ctrl{-1} & \gate{R_y(\theta_{3,L+1})}  & \gate{R_z(\theta_{6,L+1})}
    \end{quantikz}
    \end{adjustbox}

\caption{Quantum circuit used to simulate the 1D Ising model with chain length $n=3$. $L$ stands for the number of layers, while $l=1,2, ... ,L$. Each layer contains a set of parameterized $R_yR_z$ gates followed by a ladder of $CZ$ gates. The circuit ends with a final set of parameterized $R_yR_z$ gates.  The total number of parameters $N_\theta$ scales with $2n(L+1)$.}
\label{ansatz}
\end{figure}
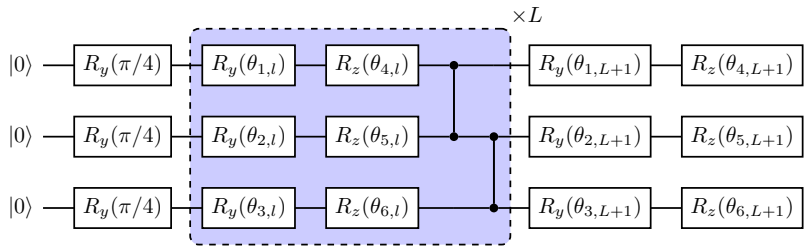

\subsection*{Quantum Circuit}
If we call $\ket{\psi(\vec{\theta})}$ the state generated by the ansatz, the optimization problem is defined as finding the values of $\vec{\theta}$ that minimize
\begin{equation}
E(\vec{\theta})=\bra{\psi(\vec{\theta})}H \ket{\psi(\vec{\theta})}\equiv E_{\theta}
\end{equation}
and the optimization is successful if  $min(E_{\theta}) = E_0$, being $E_0$ the energy of the ground state. To find the eigenvector that gives the ground state of the system, a hardware-efficient ansatz with an entanglement between adjacent qubits is expressive enough \cite{Kandala2017}. In our case, it consists of $L$ layers of parameterized $R_y R_z$ gates per qubit, followed by a ladder of controlled Z gates (CZ) plus a final parameterized layer of $R_y R_z$ gates (Fig.\;\ref{ansatz}). So, the total number of parameters $N_\theta$ is $2n(L+1)$. Finally, our quantum circuit also has  an initial layer of $R_y(\pi/4)$ gates to avoid starting directly in the state $\ket{0}^{\otimes n}$. However, this layer is not necessary unless we initialize parameters near zero. 

It is important to note that this is not the \textit{smallest} possible ansatz. For instance, a QAOA ansatz is also expressive and uses fewer parameters, as in Ref.\;\cite{PhysRevResearch.2.043246}, where only $N_\theta=n$ parameters are necessary to solve the Transverse Field Ising Model (TFIM).  This gives a more trainable quantum circuit which performs a more efficient energy minimization. This also occurs for adaptive ansatzes, which help to mitigate problems in the optimization landscape \cite{Grimsley2023,liu2023training}. Therefore, combining a highly trainable and expressive ansatz with an efficient optimization method is crucial if we want to scale towards larger systems. However, to see the capability of optimizers to avoid local minima  and make a representative statistical sample, an expressive ansatz with relatively low trainability is more suitable to work  with classical computational resources. 

\subsection*{Simulation Details}

All simulations use a locally developed code
that can be found at \url{https://gitlab.com/proyectos-cesga/quantum/react-eu/vqe_ising_chain_de}. Software version information is also available at this repository. In this work, we do not perform any technical modification in the underlying mechanism behind each optimizer. All simulations use the optimizers available from Qiskit and Scipy packages. For local optimizers, the maximum number of iterations and/or function evaluations are the unique adjusted parameters (Table \ref{table}). All optimizations using these methods finished within the specified number of iterations except some using SPSA and COBYLA ($L=4$), which run out all. For these cases optimizations end near the GS or some excited state, 	although with less precision. For DE, we fix the maximum number of iterations, the crossover strategy (\textit{bin}/\textit{exp}), and the initialization. DE simulations use \textit{Scipy} together with \textit{Multiprocessing} to run in multiple processors.

\begin{table}[h]
\centering
\begin{tabular}{|c|c|c|c|c|c|c|}
\hline
         & COBYLA            & SLSQP             & L-BFGS-B & SPSA        & DE (\textit{bin})            & DE (\textit{exp})                        \\ \hline
$N^{max}_{it}$ & 10$^5$            & 10$^3$            & 10$^4$   & 300$nL$ & 10$^5$              & 2.5 $\cdot$ 10$^4$ \\ \hline
$N^{max}_{f}$  & $N^{max}_{it}$ & $N_\theta N^{max}_{it}$ & 10$^3$ $N_\theta$ & $2N^{max}_{it}$   & $p N_\theta N^{max}_{it}$ & $p N_\theta N^{max}_{it}$             \\ \hline
\end{tabular}
\caption{Maximum number of iterations ($N^{max}_{it}$) and maximum number of function evaluations ($N^{max}_{f}$) for each optimizer used to minimize the 1D Ising model. $p$ is an integer that fixes the number of individuals per parameter when using DE.}
\label{table}
\end{table}

\section{Results}

\begin{figure}[t]
\centering
\includegraphics[scale=0.7]{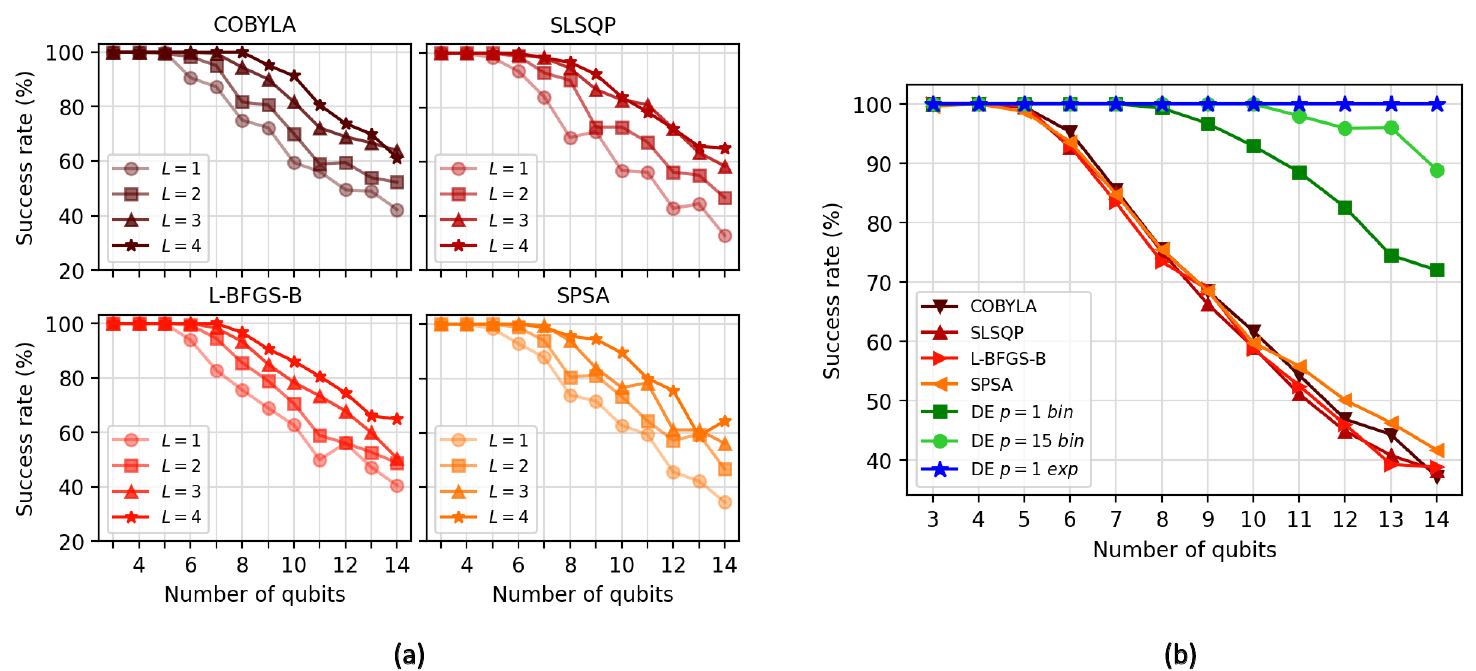}
\caption{\textbf{(a)} Success rates for 180 optimizations with different random initializations of 3-to-14 qubit circuits and a tolerance (threshold) of $10^{-2}$ for success. Each of the four curves stands for a different number of layers used in the ansatz, as shown in legends, with each subplot for an optimizer.  \textbf{(b)} Success rate ( $\delta \leq 10^{-2}$) for local optimizers and DE. Quantum circuits have $L=1$ and $N_\theta = 4n$ parameters. In optimizations with DE, $p$ indicates the number of individuals per parameter. So, the total number of individuals in the population is $p \cdot N_\theta$. Labels \textit{bin} and \textit{exp} stand for binomial and exponential crossover strategies, respectively.  $N_{opt}=1000$ in all cases except for DE with $p$=15 with a binomial crossover and DE with $p$=1 with exponential recombination, where $N_{opt}=100$.}
\label{all_optimizers}
\end{figure}

\subsection{Local optimizers: SLSQP, COBYLA, L-BFGS-B and SPSA}

First, we study the response of different used local optimizers in VQAs, which are SLSQP, COBYLA, L-BFGS-B and SPSA. These are some of the most used gradient-based and gradient-free optimizers in the VQAs literature \cite{Kandala2017,PhysRevResearch.4.013165,PRXQuantum.3.010309,McCaskey2019,Sherbert2022,Romero_2019}. However, they are predicted not to avoid barren plateaus, and in some problems, they start to get trapped in local minima as we increase the circuit complexity \cite{Arrasmith2021effectofbarren,PhysRevLett.127.120502,PhysRevResearch.2.043246}. Due to the considerable amount of optimizations, we select the success rate (SR) as our metric to compare the performance of the different methods. SR is the percentage of optimizations that finish in the ground state related to the total number of optimizations for each case. For this measurement, we set a tolerance $\delta=1 -|E_\theta/E_0|\leq 10^{-2}$ to declare the optimization as successful. For each optimization, we initialize  each parameter $\theta_{k,l}$ with a random number uniformly distributed along the interval $[-\pi,\pi)$ as they correspond to rotations in the Bloch sphere. However, we do not constrain the parameters in that range during minimization. 

Figure \ref{all_optimizers}a shows the results for each local optimizer, where each point represents a total of  180 different optimizations $N_{opt}$. Noticeably, for $L=1$, we observe that the success rate significantly decays as we increase the chain length. In particular, for chains with more than five spins, these local optimizers do not guarantee an appropriate energy minimization without exploring modifications to their algorithms. The trend improves as we increase the number of layers in the ansatz. This occurs at the expense of increasing the number of parameters and, hence, the computational time. Nevertheless, this scaling does not seem to compensate as we still have a low success rate, and the downtrend remains unalterable. 

As we will see later on, it is important to remark that any time that optimization does not find the GS, it finishes in one of the excited states of the system. This allows us to interpret it as a robust local minima problem. However, as one increases the system size, introduces noise, or increases the system entanglement, other issues, such as barren plateaus, are predicted to appear \cite{Arrasmith2021effectofbarren}, and one can deal with a variational state whose associated energy could not identify with an eigenvalue of the Hamiltonian. 

\subsection{Differential Evolution}
A possible way to avoid this local minima problem is with a multiparticle strategy that can update the \textit{particles} parameters with the remaining population members. This occurs regardless of whether the particle gets trapped in a local minimum or a barren plateau. In this work, we focus on Differential Evolution (DE), an evolutionary algorithm based on population breeding that is easy to implement and run in parallel processors \cite{Storn1997,1331145}. Together with DE, there are other evolutionary algorithms, such as the Particle Swarm Optimization, or different variations of these methods \cite{simon2013evolutionary,9046849,DENG2021107080}. However, up to now, their use is still limited in the field of quantum computing \cite{PhysRevA.105.052414,Chen_2022,Robert2021}.

In general terms, DE starts from an initial population with size $P$ instead of a single individual, as in the previously analyzed optimizers. Once the population has been created, either explicitly or randomly, DE evaluates the objective function, in our case the energy, to determine its value for the different population members and selects the one with the lowest value (\textit{best}). Individuals are vectors $\vec{x}\equiv (\theta_1, \theta_2, ..., \theta_{N_{\theta}})$ with $N_{\theta}$ elements. Remember that $N_{\theta}=2n(L+1)$ for our ansatz. From here, it starts a process in which it is defined a new candidate (\textit{mutant}) for each of the population members (\textit{target}). There are several strategies to do so \cite{AHMAD20223831}. Commonly, these strategies will employ three (or more) members of the population to generate a new \textit{mutant:} 
\begin{equation}
    \vec{x}_{mutant}=\vec{x}_{0}+F(\vec{x}_1 - \vec{x}_2)
    \label{eq:mutation}
\end{equation}
which can be all randomly chosen or using $\vec{x}_{best}$ in some of the terms. For our simulations, each \textit{mutant} is built from three vectors as in \eqref{eq:mutation} but using $\vec{x}_0=\vec{x}_{best}$.  $F$ is the mutation factor. Once we have a candidate for each population member, a recombination phase between the targets and mutants starts. In there, some parameters of the target are replaced by the mutant\textquotesingle s ones following a crossbreeding strategy. This strategy can be binomial, where each parameter has a probability $\mathcal{C}$ of being changed by the mutant\textquotesingle s one, or exponential, where all parameters between two randomly chosen elements of the target are replaced \cite{exponential}. The process finishes with the evaluation of the energy of the modified target vector. The modified target replaces the original target if the energy is lower or is discarded if it is higher. This process is repeated until convergence
\begin{equation}
\sqrt{\frac{\sum_k^P(E_k-\bar{E})^2}{P}} \leq t + t' |\bar{E}|
\end{equation}
where $t$, and $ t'$ are the absolute and relative tolerances, and $\bar{E}$ is the average energy of the population.

With this in mind, we perform our simulations using DE. For this task, we use the \textit{scipy.optimize} package where DE is already implemented, allowing direct parallelization in different workers using \textit{Multiprocessing}. In this case, we increase $N_{opt}$ up to 1000 for the local optimizers, repeating the optimizations for $L=1$ to make our comparison with DE more reliable. For DE, we choose a random initialization with \textit{halton}  (method provided by \textit{Scipy}), which allows maximizing the parameter space explored at the beginning \cite{Halton}. In this way, the population size $P$ is exactly  the product of $p \cdot N_{\theta}$, being $p$ an integer that fixes the number of individuals per parameter.  We execute two sets of simulations: one using $p\in\{1,15\}$ with a binomial crossover and a second with $p=1$ and an exponential recombination strategy (Fig. \ref{all_optimizers}b). In these cases, $N_{opt}=100$ except for DE with binomial crossover and $p=1$ where $N_{opt}=1000$. It is important to remark that simulations with $p=1$ allow a direct comparison between DE and the previously analyzed methods. SLSQP and L-BFGS-B (SPSA) compute (approximates) the gradients of $N_{\theta}$ components, and COBYLA computes $N_{\theta}$ distances. Therefore, for these optimizers, the total number of \textit{virtual targets} is $P=N_{\theta}$. Tolerance values for the stopping criteria are $t=0$ and $t'=10^{-5}$.

\begin{figure}[t]
\centering
\includegraphics[scale=0.75]{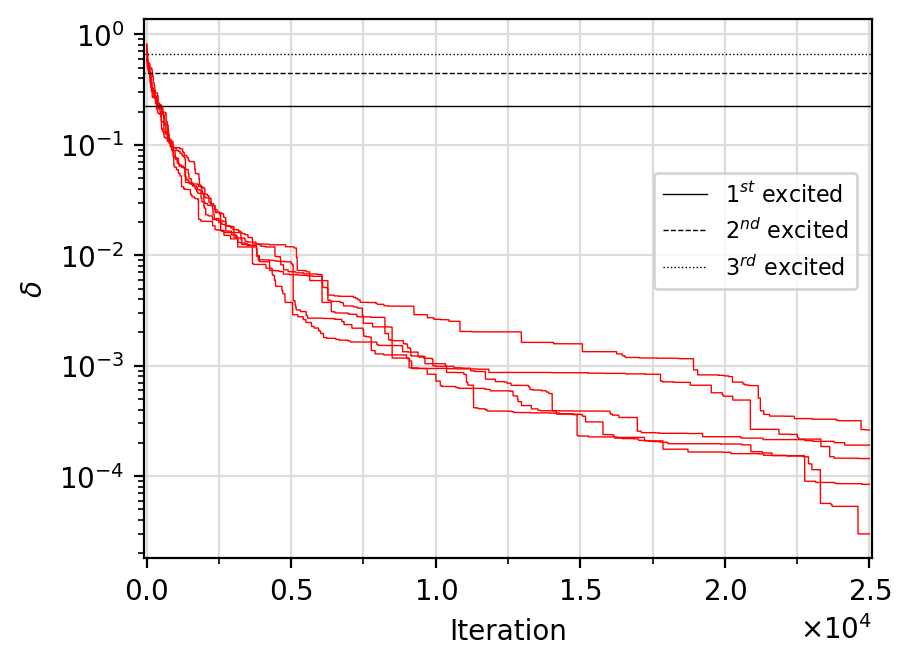}
\caption{Relative error $\delta=1 -|E_\theta/E_0|$ vs. iteration for a 10 qubits Ising chain without magnetic field using DE with exponential crossover ($p=1$). The plot shows five different random initialization of the population $P$. Optimizations leap all excited states in the first 1000 iterations approximately. However, they need much more iterations to improve the convergence to the GS. }
\label{cost}
\end{figure}

\subsubsection*{Binomial crossover}

For simulations with a binomial crossover, we observe that even in the same conditions ($p=1$), DE clearly outperforms local optimizers. This is because, even if one target gets trapped in a local minimum, DE can continue evolving. Therefore, this individual has a non-zero probability of eluding the local minimum, thanks to the remaining members. This is not possible for the previous methods unless we explore modifications to their mechanism. Nevertheless, as we increase the number of qubits, the number of possible local minima increases exponentially as $2^n$ while the ground state always maintains the same degeneracy. It stands to reason that although DE offers better results than the previous methods, even when the ansatz contains more than a single layer, the SR also decays, following, in general terms, the same trend as the previous methods but for a higher number of qubits. Based on this knowledge, increasing the population size $P$ is one way to enhance the SR using DE.  To this aim, we execute another set of simulations with   $p=15$,  for which we find that the SR always holds above $88\%$. This configuration improves the previous results and opens the path to further enhance the SR by only increasing $P$.  However, this will severely enlarge the number of function evaluations needed for the optimization, which is something to look out for when executing in a quantum computer. Notice that DE evaluates $P=p\cdot N_\theta$ circuits in each iteration in contrast to SLSQP and L-BFGS-B, which only require $N_\theta$, and COBYLA, SPSA optimizers, which make one and two function evaluations per iteration, respectively. Then, it is convenient to explore other alternatives rather than directly increasing $P$, for instance, modifying the recombination scheme.

\subsubsection*{Exponential crossover}

Despite the significant improvement, using a binomial crossover together with the mutation scheme given by \eqref{eq:mutation} for $p\in\{1,15\}$ does not guarantee the avoidance of all local minima in the range studied. This suggests that the configuration used can still cause all individuals to cluster in a local minimum if $\vec{x}_{best}$ and other targets get trapped. 

To try to avoid this, we change the recombination criteria to exponential, which results in a more aggressive mutation scheme. An exponential crossover drastically modifies the target, maximizing the  parameter space exploration and the probability of \textit{tunneling} in a minimum. However, we lose convergence when approaching the global minimum, given that $\vec{x}_{best}$ has less influence, and most of the modified targets will not improve the energy. In this way, we fix the maximum number of iterations allowed to $2.5 \cdot 10^4$, which is far from the number of iterations $N_{it}$ done in the previous simulations that encompass from a few hundred to around $8\cdot10^3$ for $p=1$ and 14 qubits. We can see in Figure \ref{all_optimizers}  how this DE configuration reaches the GS with a 100\% probability in the range studied.

\begin{figure*}[t]
    \includegraphics[scale=0.75]{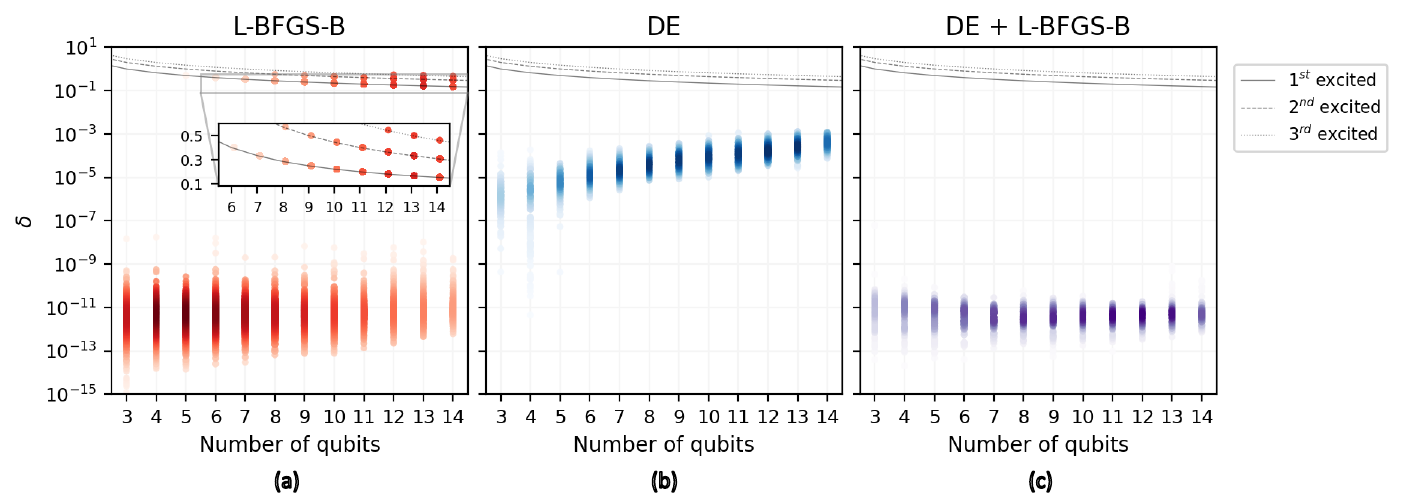}
    \caption{Relative errors comparison using local, global, and hybrid optimization strategies. \textbf{(a)} L-BFGS-B data of $N_{opt} = 1000$ experiments. The inset shows the excited states (grey curves) found during optimizations. \textbf{(b)} Results using DE with $p$=1 and exponential crossover for $N_{opt} = 100$. Relative error grows with the number of qubits given that optimizations are limited by a constant number of iterations ($N_{it}^{max}=2.5 \cdot 10^4$). \textbf{(c)} Results of a hybrid strategy, where we repeat the previous DE simulations to avoid convergence to excited states, but we subsequently call the L-BFGS-B optimizer to polish $\vec{x}_{best}$.}
    \label{deltas}
\end{figure*}

However, this stands when the upper limit of $\delta$ that defines SR is $10^{-2}$. More precise approaches to the GS ($\delta<< 10^{-2}$) will require a larger number of iterations to maintain the SR (Figure \ref{cost}). Nevertheless, this is not associated with local minima, as in the previous cases, but with convergence. Therefore, although the relative error achieved is enough in many cases and lower than feasible VQE errors in current quantum hardware, our next goal is to minimize $N_{it}$ and, if possible, reach a $\delta$ similar to what gradient methods offer when successfully finding the GS (Figure \ref{deltas}a) \cite{PhysRevResearch.2.043246}. In Figure \ref{deltas}, we can see the complete set of final relative errors using the L-BFGS-B ($N_{opt}=10^3$) method and DE ($p=1$, \textit{exp}, $N_{opt}=10^2$). We can see how the L-BFGS-B method increasingly finds more excited states of $H$ as we grow in the number of qubits. On the contrary, DE always avoids these states but features a relative error higher than the L-BFGS-B. The causes are that we limited the maximum number of iterations to $N_{it}^{max}=2.5\cdot 10^4$ and that the exponential crossover leads to slow convergence.

\subsection{Hybrid optimization}

The straightforward strategy to reduce $\delta$ is using a gradient-based optimizer after DE with exponential crossover. In this case, we call the L-BFGS-B method, given that it is the gradient-based optimizer with  the lowest tolerance value ($10^{-15}$) by default. Our simulations ($N_{opt}=100$) take the individual with the lower energy $\vec{x}_{best}$ after $2.5 \cdot 10^4$ iterations and initialize a new optimization with the L-BFGS-B method from its parameters. Results are shown in Figure \ref{deltas}c, where this hybrid optimization avoids all local minima and finds the GS with high accuracy.

On the other hand, reducing $N_{it}$, and hence the number of circuit executions can be done efficiently by calling earlier a gradient-based optimizer. However,  this could imply knowing a priori some information from your energy spectra, such as the energy difference between the GS and the first excited level. Another possibility is to alternate the use of DE with \textit{exp} and \textit{bin} recombination schemes.

\section{Towards strongly correlated systems}
In this final section, we apply the same methodology to a more complex model from the physical point of view, the Hubbard model. Although the 1D Ising model can work as a complex optimization problem, as we saw in the previous section employing a quantum circuit that is agnostic to the system, its spin configurations can be analytically obtained. In fact, going to the base of fermionic operators by means of an inverse Jordan-Wigner transformation, the Ising model studied just involves quadratic terms of fermionic creation/annihilation operators, which are classically tractable. This is not the case for the exact calculation of strongly correlated systems, which include in their Hamiltonians higher-order terms, and for what quantum computing holds as a promising tool in diverse fields. To delve into correlations we use the most common alternative, which is the Hubbard model in a 1D lattice \cite{PhysRevResearch.4.013165}
\begin{equation}
H= H_t + H_U=- t \sum_s\sum_{i}^n  \left(a^\dagger_{i,s} a_{i+1,s} + a^\dagger_{i+1,s} a_{i,s}  \right) + U \sum_i^n  a^\dagger_{i,\uparrow} a_{i,\uparrow} a^\dagger_{i,\downarrow} a_{i,\downarrow}
\label{Hubbard}
\end{equation}
where the index $s \in \lbrace \uparrow, \downarrow \rbrace$ denotes the spin of the electrons, $t$ is the hopping amplitude that we assume constant for simplicity, and U is the Hubbard constant that represents the strength of the on-site Coulomb repulsion. The operator $a^\dagger_{i,s}$ ($a_{i,s}$) creates (annihilates) an electron of spin $s$ at site $i$. We take periodic boundary conditions. In this case, to see DE performance minimizing the energy, we do not consider a quantum circuit agnostic to the system to ensure expressibility. By contrast, we use a Hamiltonian Variational Ansatz (HVA) that is adequate to model a wide range of condensed matter systems \cite{PRXQuantum.1.020319,PhysRevResearch.4.023190,PhysRevA.92.042303}. The HVA is a QAOA-inspired ansatz based on adiabatic evolution to achieve the ground state of the system \cite{PRXQuantum.1.020319}. It takes the Hamiltonian to construct the quantum circuit, so we need to map it into Pauli operators using, for instance, a Jordan-Wigner transformation \cite{TILLY20221}.  The resulting Hamiltonian is expressed then as a sum of commuting groups, i.e., $H=\sum_k H_k$ where $[H_k, H_{k'}]\neq 0$, that determines our quantum circuit. The HVA generates a trial wavefunction of the form 

\begin{figure*}[t]
    \includegraphics[scale=0.95]{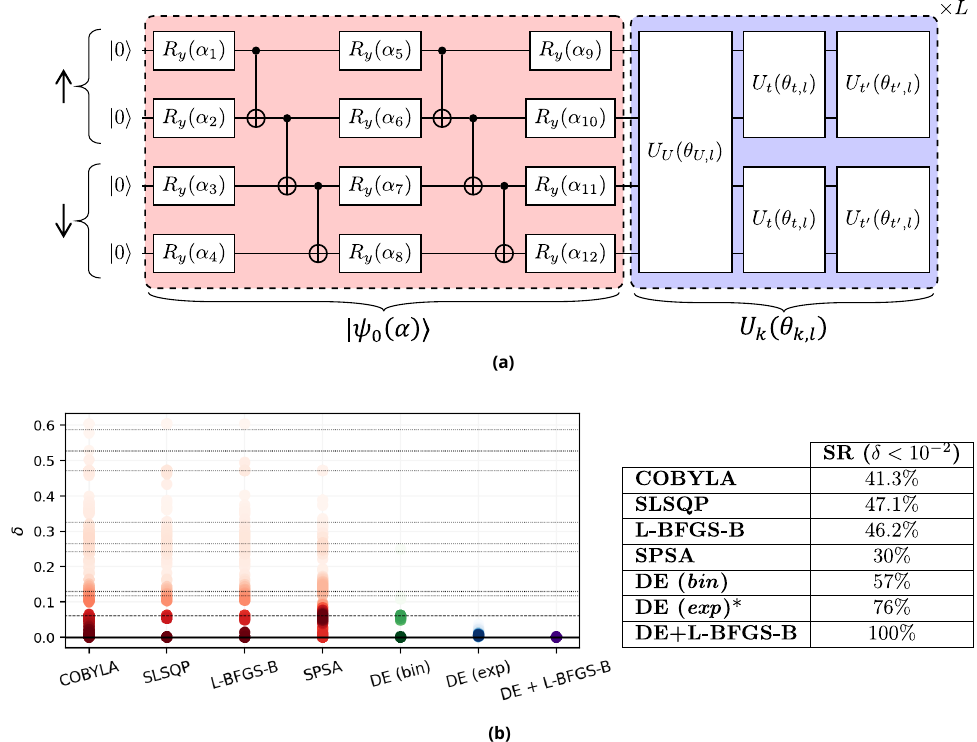}
    \caption{\textbf{(a)} Schematic illustration of the parameterized quantum circuit used to find the ground state of the Hubbard Hamiltonian \eqref{Hubbard}. The figure exemplifies the circuit used to solve a two-site lattice ($n=4$ qubits). The part colored in red represents the parameterized initial state belonging to the non-interacting term $H_t$.  The blue one corresponds to the layers of the HVA. The unitary operator $U_U(\theta)=e^{-i \theta H_U}$ is associated with the Hubbard term in \eqref{Hubbard}, while $U_t$ and $U'_t$ come from the two sets of commuting groups defined from the Jordan-Wigner transformation of $H_t$. \textbf{(b)} Plot of relative error comparison using several optimizers in a 8-qubit 1D Hubbard model. We include a table to show the SR for each of them. The solid, dashed, and dotted lines illustrate the ground state, the first excited, and the upper energy levels, respectively. All optimizations using DE with exponential crossover strategy stopped at $N_{it}^{max}=2.5 \cdot 10^4$ iterations. Therefore, its SR is not associated with local minima but limited  to $N_{it}^{max}$.  Simulations using the hybrid optimization scheme confirm this fact featuring a 100\% SR.}
    \label{Hubbard figure}
\end{figure*}

\begin{equation}
\ket{\Psi({\theta})}=\prod_l^L \left( \prod_k e^{-i\theta_{k,l} H_k} \right) \ket{\psi_0}
\label{HVAwavefunction}
\end{equation}
where each layer $l$ is a product of unitary operators $U_{k}(\theta_{k,l})=e^{-i\theta_{k,l} H_k}$ that can be translated to qubit instructions, and $\ket{\psi_0}$ is the ground state of some part of the Hamiltonian, for instance, $H_t$. There are different ways to obtain $\ket{\psi_0}$ with variable fidelity with respect to the target state \cite{PhysRevResearch.4.023190, PhysRevB.103.L241113}. However, in contrast to the general prescription, we opt to make the preparation of $\ket{\psi_0}$ part of our variational protocol up to some extent. We do this to increment the complexity of the optimization procedure as the HVA is an expressible and trainable ansatz. Otherwise, we would need to increase the number of qubits to see the effects of a complex optimization landscape, as it happens for the TFIM with the QAOA ansatz \cite{PhysRevResearch.2.043246}. As a consequence, we would require a considerable amount of cores and memory for a classical simulation. To ensure expressibility, we prepare a parameterized wavefunction inspired in the ground state of the non-interacting part of the Hubbard Hamiltonian \eqref{Hubbard}, i.e., the one consisting of quadratic terms. We perform a classical calculation of the transformation matrix that allows us to diagonalize the non-interacting part of \eqref{Hubbard}. Then, we compute the Slater determinant associated with the minimum eigenvalue. In the computational basis, this state $\ket{\psi_0}$ can be expressed as a product of NOT gates (depending on the number of electrons),  single qubit rotations $R_y R_z$, and CNOT gates \cite{PhysRevB.103.L241113}. Thus, we can build the parameterized initial state $\ket{\psi_0}$ with the same structure as the layers of the hardware efficient ansatz used for the 1D Ising model but removing the parameterized $R_z$ gates and substituting the CZ gates with CNOTs. We can disregard $R_z$ gates since the Hamiltonian $H_t$ is real and, therefore, also its eigenstates. The structure of the full quantum circuit is shown in Figure \ref{Hubbard figure}a. For simulations, we focus on a four-site Hubbard model ($t=1$, $U=1$), for which we need a total of $n=8$ qubits due to the electron\textquotesingle s spin. Our quantum circuit takes two layers for the variational initial state and $n/2$ layers for the HVA, i.e., $N_{\theta}=36$ parameters. We perform again several sets of simulations using i) all previous local optimizers, ii) DE with $p=1$ and binomial crossover, iii) DE with $p=1$ and exponential crossover and iv) a hybrid optimization scheme DE (\textit{exp})/L-BFGS-B as the one previously employed. We execute $N_{opt}=1000$ optimizations with random initialization for the local optimizers and $N_{opt}=100$ for the ones using DE. The parameters of the optimizers remain unchanged except for SPSA, for which we need to enlarge $N_{it}^{max}$ to $5 \cdot 10^4$.  

As shown in Figure \ref{Hubbard figure}b, local optimizers fail to minimize the energy featuring a relatively low SR. As before, DE with binomial crossover provides better results than these algorithms but still gives a substantial amount of incorrect outputs. For its part, DE evolution with exponential crossover approaches quite well to the GS energy after $N_{it}=2.5\cdot 10^4$ iterations. That is an outstanding point, given that any of the optimizations end up meeting the convergence criteria. Therefore, the success rate obtained is only a convergence problem associated with limiting $N_{it}$ and our tolerance threshold for $\delta$. In this way, we suppose that DE \textit{exp} has avoided all traps in the optimization landscape. This is demonstrated in the final set of simulations, where the L-BFGS-B optimizer applied to the best candidate of DE \textit{exp} achieves the ground state energy with 100\% probability.  Finally, another important point comes from the fact that, for local optimizers and DE \textit{bin}, optimizations stop at energy values that are not eigenvalues of the Hamiltonian. This could be associated with other traps in the optimization landscape, like barren plateaus, rather than local minima. It stands to reason that these areas appear due to the complexity of the problem, the variational initial state, and the breakout of adiabaticity caused by the random initialization of the parameters \cite{PhysRevResearch.2.043246}. Of course, this does not harm the ansatz expressibility, just the complexity of the optimization. Despite that, DE with exponential crossover avoids all conflicting points for the case studied.

\section*{Discussion}

In the near term, the most powerful applications of quantum computers rely on VQAs. However, their applicability to large systems is a current challenge due to optimization-landscape problems. In this context, the ansatz selection, to enhance trainability maintaining expressibility, and the optimization method, are crucial for their performance. In this work, we show DE as a strong candidate to lead the optimization in VQAs. In particular, using basic DE configurations, our results evidence the ability of DE to avoid local minima, clearly outperforming four of the most used optimizers in VQAs literature. 

We find that DE with binomial crossover performs better than local optimizers and provides strong convergence. However, it fails to avoid all local minima. DE with exponential crossover features a more aggressive mutation scheme, increases the parameter space exploration and the probability of eluding excited states, but needs a large $N_{it}$ to maintain the same accuracy as the other methods when finding the ground state. To avoid local minima, we can apply other alternatives instead of or together with the exponential crossover. For instance, increasing $P$, using more individuals in the mutation scheme given by \eqref{eq:mutation}, or using a random individual for $\vec{x}_0$ instead of $\vec{x}_{best}$. Nevertheless, we can not expect these variations to need fewer circuit executions to reach the ground state. In this context, hybrid optimization schemes such as the one employed or another that intersperse DE with a local optimizer  seem to be more correct approaches to this purpose, as well as to elude local minima.     

Besides their simplicity, the most relevant characteristic of DE resides in its ability to continue evolving regardless of one or several individuals getting stuck, so it could be immune/resilient to vanishing gradient areas. This comes reasonably from Figure \ref{Hubbard figure}b, although more experiments  should address it specifically. In addition, DE could present additional benefits against gradient-based algorithms since computing gradients requires more shots as they become small \cite{Romero_2019}.  

This work uses exact simulations to find the GS in a VQE to see the capability of the different optimizers to avoid local minima in ideal conditions. In this context, DE and hybrid optimizations with DE demonstrate a compelling advantage to scale VQAs. Future work addressing DE performance in noisy environments and real quantum computers can reaffirm DE-based optimization as a robust candidate to walk toward the so-called \textit{quantum advantage} in VQAs.

\section*{Data availability}

All data generated or analysed during this study are available  from the corresponding author upon reasonable request.

\section{Acknowledgments}
We thank the CESGA Quantum Computing group members for feedback and the stimulating intellectual environment they provide. We thank Lois Orosa, Prof. Javier Mas, and Juan Santos for feedback and helpful discussions. This work was supported by Axencia Galega de Innovación through the Grant Agreement ``Despregamento dunha infraestrutura baseada en tecnoloxías cuánticas da información que permita impulsar a I+D+I en Galicia'' within the program FEDER Galicia 2014-2020. A. Gómez was supported by MICIN through the European Union NextGenerationEU recovery plan (PRTR-C17.I1), and by the Galician Regional Government through the “Planes Complementarios de I+D+I con las Comunidades Autónomas” in Quantum Communication. Simulations on this work were performed using the Finisterrae III Supercomputer, funded by the project CESGA-01 FINISTERRAE III.

\nocite{*}
\bibstyle{naturemag-doi}

\providecommand{\noopsort}[1]{}\providecommand{\singleletter}[1]{#1}%

\end{document}